\documentclass[12pt]{article}
\usepackage{amssymb}
\input{epsf}
\newcommand{\ssy}[5]{#1, {#2} {\bf #3}, #5 (#4)}
\newcommand{\sstar}{\;{\star}\;}
\newcommand{\kart}[4]{\begin{figure}[#1]\begin{center}\leavevmode%
\epsfysize=#2\epsffile{#3.ps}\end{center}\caption{#4}\end{figure}}
\newcommand{\Bd}{\text{Bd}\,}
\newcommand{\text}[1]{\mbox{\rm #1}}
\newcommand{\Vn}{\text{Int}\,}
\newcommand{\JS}[1]{\ensuremath{J^+(S_{#1})}}
\newcommand{\BS}[1]{(#1)_{\mathbb J}}
\author{S. V. Krasnikov\thanks{Email: \it redish@pulkovo.spb.su}\\
The Central\\
Astronomical Observatory at Pulkovo, St Petersburg,\\
196140, Russia}
\title{Hyperfast interstellar travel in general relativity}
\date{}
\begin{document}
\maketitle

\begin{abstract}
The problem is discussed of whether a traveller can reach a remote
object and return back sooner than a photon would  when taken into
account that 
the traveller can partly control the geometry of his world. It is
argued that 
under some reasonable assumptions in globally hyperbolic spacetimes
the traveller cannot hasten reaching the destination.  Nevertheless, it is
perhaps possible for him to make an arbitrarily long  \emph{round-trip}
within an arbitrarily short (from the point of view of a terrestrial
observer) time.
\end{abstract}

\section{Introduction}                 

Everybody knows that nothing can move faster than light. The
regrettable consequences of this fact are also well known. Most of
the interesting or promising in possible colonization objects are so
distant from us that the light barrier seems to make an
insurmountable obstacle for any expedition. It is, for example,
$200\,$pc from us to the Polar star, $500\,$pc to Deneb and $\sim
10\,$kpc to the centre of the Galaxy, not to mention other galaxies
(hundreds of kiloparsecs). It makes no sense to send an expedition if
we know that thousands of years will elapse before we receive its
report.\footnote{The dismal fate of an astronaut returning to the
absolutely new (and alien to him) world was described in many science
fiction stories, e.~g.~in \cite{Lem}.} On the other hand, the
prospects of being confined forever to the Solar system without any
hope of visiting other civilizations or examining closely black
holes, supergiants, and other marvels are so gloomy that it seems
necessary to search for some way out.
\par
In the present paper we consider this problem in the context of
general relativity. Of course the light barrier exists here too. The
point, however, is that in GR one can try to change the time
necessary for some travel not only by varying one's speed but also,
as we shall show, by changing the distance one is to cover.
\par
To put the question more specific, assume that we emit a beam of test
particles from the Earth to Deneb (the event $S$). The particles move
with all possible (sub)luminal speeds and by definition do not exert
any effect on the surrounding world.  The beam reaches Deneb (with
the arrival time of the first particle $t_{\mathcal{D}}$ by Deneb's
clocks), reflects there from something, and returns to the Earth.
Denote by $\Delta \tau_\mathcal E$ ($\tau_\mathcal E$ is the Earth's
proper time) the time interval between $S$ and the return of the
first particle (the event $R$). The problem of interstellar travel
lies just in the large typical $\Delta \tau_\mathcal E$. It is
conceivable of course that a particle will meet a traversible
wormhole leading to Deneb or an appropriate distortion of space
shortening its way (see \cite{Alc} and Example~4 below), but one
cannot hope to meet such a convenient wormhole each time one wants to
travel (unless one makes them oneself, which is impossible for the
{\em test} particles).  Suppose now that instead of emitting the test
particles we launch a spaceship (i.~e.~something that does act on the
surrounding space) in $S$.  Then the question we discuss in this
paper can be formulated as follows:
\begin{quote}
\em Is it possible
that the spaceship will reach Deneb and then return to the Earth in
$\Delta \tau'<\Delta \tau_\mathcal E$?
\end{quote}
By ``possible" we mean ``possible, at least in principle, from
the causal point of view". The use of tachyons, for example, enables,
as is shown in \cite{Alc}, even a nontachionic spaceship to hasten its
arrival.  Suppose, however, that tachyons are forbidden (as well as
all other means for changing the metric with violating what we call
below ``utter causality'').  The main result of the paper is the
demonstration of the fact that even under this condition the answer
to the above question is positive. Moreover, in some cases (when
global hyperbolicity is violated) even $t_{\mathcal{D}}$ can be
lessened.

\section{Causal changes}
\subsection{Changes of spacetime}
In this section we make the question posed in the Introduction
more concrete.  As the point at issue is the effects caused by
\emph{modifying the (four-dimensional) world}  (that is, by changing its
metric or even topology), one may immediately ask, Modifying from
what? To clarify this point first note that though we treat the
geometry of the world classically throughout the paper (that is, we
describe the world by a spacetime, i.~e., by a smooth
Lorentzian connected globally inextendible Hausdorff manifold),
no special restrictions are
imposed for a while on  matter fields (and thus on the right-hand side of the
Einstein equations).  In particular,
{\em
	it is not implied that the
	matter fields (or particles) obey any specific classical
	differential equations.}
\par

Now consider an experiment with two possible results.
\par

\noindent{\bf Example 1.} A device set on a spaceship first
polarizes an electron in the $y$-direction 
and then measures the $x$-component of its spin $\sigma_x$. If the
result of the measurement is $\sigma_x=+1/2$, the device turns the
spaceship to the right; otherwise it does not.
\par

\noindent{\bf Example 2.} A device set on a (very massive) spaceship
tosses a coin. If it falls on the reverse the device turns the spaceship
to the right; otherwise it does not.
\par

\noindent{\bf Comment.} One could argue that Example 2 is inadequate since
(due to the classical nature of the experiment) there is actually
{\em one\/} possible 
result in that experiment. That is  indeed the case. However, let us
	\begin{enumerate}
	\item[({\it i})] Assume that before being tossed the coin had never
	interacted with  anything.
	\item[({\it i})] Neglect the contribution of the coin to the metric of
	the world.
	\end{enumerate}
Of course, items ({\it i,ii}) constitute an {\em approximation} and
situations are conceivable for which such an approximation is invalid
(e.~g., item~($i$) can be illegal if the same coin was already used as a
lot in another experiment involving large masses). We do not consider such
situations. And if ({\it i,ii}) are adopted, experiment 2  can well be
considered as an experiment with two possible results (the coin, in
fact, is unobservable before the experiment). 
\par
Both situations described above suffer 
some lack of determinism (originating from the quantum indeterminism
in the first case and from coarsening the classical description in
the second). Namely, the spaceship is described
now by a body whose evolution is not fixed uniquely by the initial
data (in other words, its trajectory (non-analytic, though smooth) is
no longer a solution of any ``good" differential equation). However, as
stated above, this does not matter much.

\paragraph{}
So, depending on which result is realized in the experiment (all
other factors being the same; see below) our world
must be described by one of {\em two different spacetimes}.  It is
the comparison between these two spacetimes that we are interested in.
\paragraph{Notation.} Let $M_1$
and $M_2$ be two spacetimes with a pair of inextendible timelike curves
$\mathcal{E}_i,{\mathcal{D}}_i\subset M_i\,$ in each (throughout the
paper $i,j=1,2$).  One of these spacetimes, $M_1$ say, is to describe
our world under the assumption that we emit test particles at some
moment $S_1 \in \mathcal{E}_1$ and the other, under the assumption that
instead of the particles we launch a spaceship in $S_2$, where $S_2 \in
M_2$ corresponds in a sense (see below) to $S_1$. The curves
${\mathcal{E}}_i$ and $\mathcal{D}_i$ are the world lines of 
Earth and Deneb, respectively.  We require that the two pairs of
points exist:
	\begin{equation}
	F_i \equiv \Bd \big(J^+(S_i)\big) \cap 
	{\mathcal{D}}_i \quad \mbox{and} \quad
	R_i \equiv \Bd \big(J^+(F_i)\big) \cap {\mathcal{E}}_i.
	\label{Poi}
	\end{equation}
These points mark the restrictions posed by the light barrier in each
spacetime. Nothing moving with a subluminal speed in the world
$M_i$ can reach Deneb sooner than in $F_i$ or return to the Earth
sooner than in $R_i$. What we shall study is just the relative
positions of $S_i\,,F_i\,,R_i$ for $i=1,2$ when the difference in the
spacetimes $M_1$ and $M_2$ is of such a nature (below we formulate
the necessary geometrical criterion) that it can be completely ascribed to the
pilot's activity after $S$.

%
\subsection{``Utter causality"} The effect produced by the traveller
on spacetime need not be weak. For example, by a (relatively)
small expenditure of energy the spaceship can break the equilibrium
in some close binary system on its way, thus provoking  the collapse.
The causal structures of $M_1$ and $M_2$ in such a case will differ
radically.  If an advanced civilization (to which it is usual to
refer) will cope with topology changes, it may turn out that $M_1$
and $M_2$ are even nondiffeomorphic. So the spacetimes under discussion
may differ considerably. On the other hand, we want them to be not
{\em too\/} different:
\par

{\bf 1.} 
 The pilot of the spaceship deciding in $S$ whether or
not to fly to Deneb knows the pilot's past and in our model we would
prefer
that the pilot's decision could not change this past. This
restriction is not incompatible 
with the fact that the pilot can make different decisions (see the
preceding subsection).\par

{\bf 2.} 
The absence of tachyons (i.~e.~fields violating the postulate of local
causality \cite{HawE}), does not mean by itself that one (located in,
say, point $A$) cannot act on events lying off one's ``causal
future'' (i.~e.~off $J^+(A)$).
\begin{itemize}
\item[(\emph i)] Matter fields are conceivable that while satisfying 
local causality themselves do not provide local causality to the
metric. In other words, they afford a unique solution to the Cauchy
problem for the metric, not in $D^+({\cal P})$ (cf.~chapter 7 in
\cite{HawE}), but in some smaller region only. In the presence of
such fields the metric at a point $B$ might depend on the fields at
points outside $J^-(B)$. That is, the metric itself would act as a
tachyon field in such a case.
\item[(\emph{ii})] Let $M_1$ be the Minkowski space with coordinates
$(t_1,x_1^\mu)$ and $M_2$ be a spacetime with coordinates
$(t_2,x_2^\mu)$ and with the metric flat at the region $x^1_2>t_2$,
but nonflat otherwise (such a spacetime describes, for example,
propagation of a plane electromagnetic wave). Intuition suggests that
the difference between $M_1$ and $M_2$ is not accountable to the
activity of an observer located in the origin of the coordinates, but
neither local causality nor any other principle of general relativity
forbids such an interpretation. 
\end{itemize}
In the model we construct we want to abandon any possibility of such
``acausal'' action on the metric. In other words, we want the condition
relating $M_1$ and $M_2$ to imply that these worlds are
 the same in events that cannot be causally connected
to $S$. This requirement can be called the {\em principle of utter
causality}.


\subsection{Relating condition} In this section we formulate the
condition relating $M_1$ to $M_2$. Namely, we require that $M_i$
should ``diverge by $S$" (see below). It should be stressed that from
the logical point of view this condition is just a
{\em physical postulate.} Being concerned only with the relation between two
``possible" worlds, this new postulate does not affect any previously
known results. In defense of restrictions imposed by our postulate on
$M_i$ we can say that it does not contradict any known facts.
Moreover, in the absence of tachyons (in the broad sense, see 
item~(\emph i) above) it is hard to conceive of a mechanism violating it.
\par
Formulating the condition under discussion we would like to base it on
the ``principle of utter causality". In doing so, however,
 we meet a  circle: to
find out whether a point is causally connected to $S_i$ we must know
the metric of the spacetime $M_i$, while the metric at a
point depends in its turn on whether or not the point can be causally
connected to $S_i$.  That is why we cannot simply require that
$M_i\setminus J^+(S_i)$  be isometric. The following example
shows that this may not be the case even when utter causality
apparently holds.

\paragraph{Example 3. ``Hyper-jump."}                    
Let $M_1$ be the Minkowski plane with $S_1$ located at the origin of
the coordinates and let $M_2$ be the spacetime (similar to the
Deutsch-Politzer space) obtained from $M_1$ by the following
procedure (see Fig.~1). Two cuts are made, one along a segment $l$
lying in $I^+(S_1)$ and another along a segment $l'$ lying off
$J^+(S_1)$ and obtained from $l$ by a translation. The four points
bounding $l,\,l'$ are removed and the lower (or the left, if $l$ is
vertical) bank of each cut is glued to the upper (or to
the right) bank of the other.  Note that we can vary the metric in
the shadowed region without violating utter causality though this
region ``corresponds" to a part of $M_1\setminus J^+(S_1)$.

\par
\kart{h,t,b}{30ex}{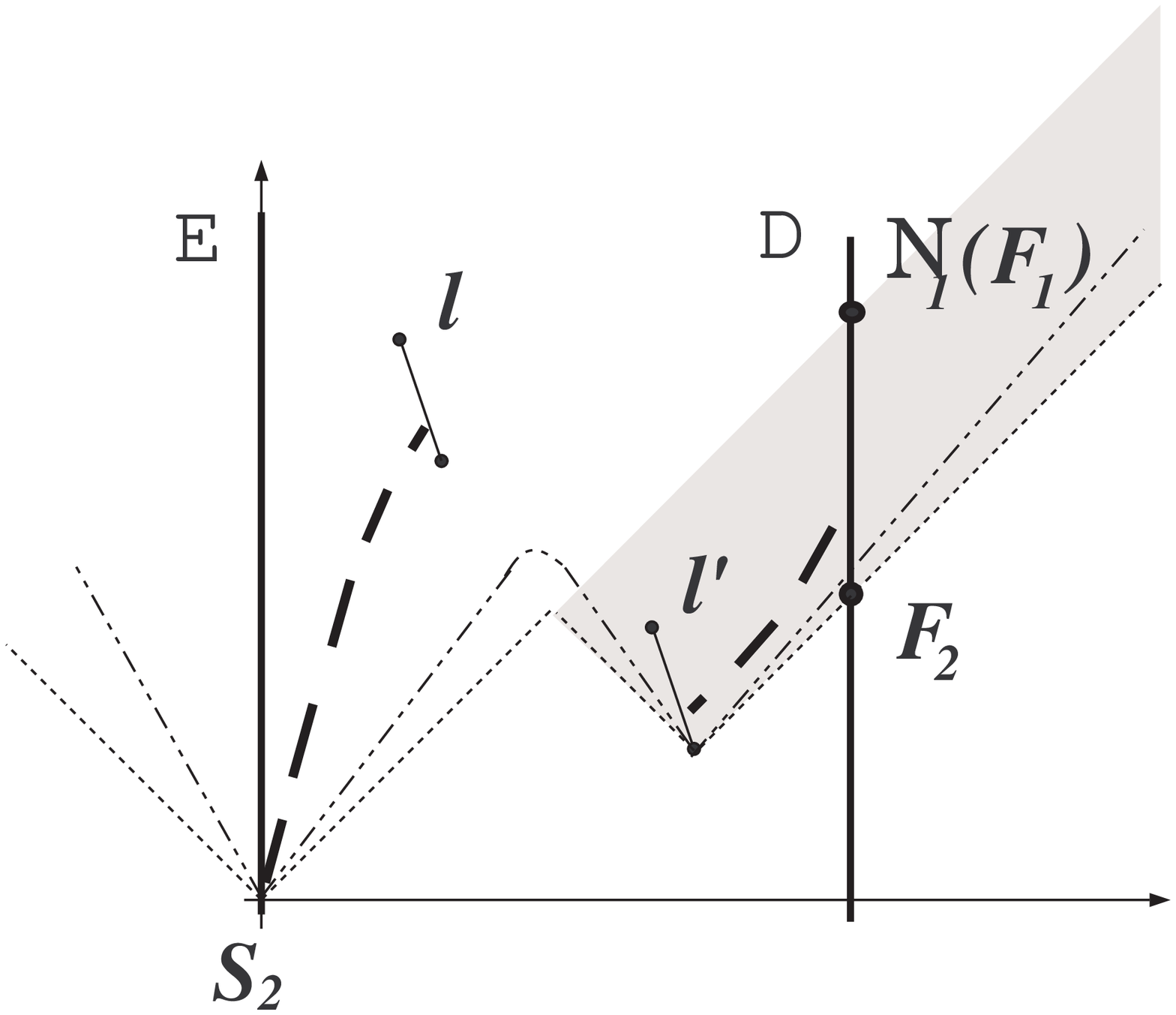}%
{``Hyper-jump." The thick dashed line depicts
an allowed  world line of a spaceship.} 
To overcome this circle we shall formulate our relating condition in
terms of the boundaries of the ``unchanged'' regions.
%
\paragraph{Notation.} Below we deal with two spacetimes $N_i \subset M_i$ 
related by an isometry $\phi\colon\; N_2=\phi(N_1)$. To
shorten notation we shall write sometimes $X_{(1)}$ for a subset $X
\subset N_1$, and $X_{(2)}$ for $\phi(X)$.  The notation $ A \sstar B$
for points $A,\,B$ will mean that there exists a sequence $\{a_n\}$:
	$$
	a_{n(i)} \to A, \quad  a_{n(j)} \to B .
	$$
Clearly if $A\in N_1$, then $ A \sstar B \neq A $  means simply
$B=\phi(A)$. Lastly, $\vphantom{\bigg|}\mathbb J \equiv
\overline{\JS{1}} \cup \overline{\JS{2}} $. 
%
%
\paragraph{Definition 1.} We  call  spacetimes $M_1,\: M_2$
{\em diverging by the event $S_1$} (or {\em by $S_2$},
or simply {\em by $S$}) if there exist
open sets $N_i\subset M_i$, points $S_i$, 
 and an
isometry $ \phi\colon\; N_1 \mapsto N_2 $ such that
 $ I^-(S_2) = \phi \big( I^-(S_1) \big) $ and
	\begin{equation}
	(Q_j \cup Q_k) \cap { \mathbb J } \neq \emptyset
	\label{Con}
	\end{equation}
whenever $Q_j \in \Bd N_j$ and $Q_j \sstar Q_k \neq Q_j$.
\paragraph{Comment.} In the example considered above the two
spacetimes  diverged by $S$. Note that \\
(\emph{i}) The possible choice of $N_i$ is not unique. The dotted
lines in Fig.~1 bound from above two different regions
that can be chosen as $N_2$.\\
(\emph{ii}) $A_{(2)} \prec B_{(2)}$ does not necessarily imply
$_{(1)} \prec B_{(1)}$.
(\emph{iii}) Points constituting the boundary of $N$ fall into two
types, some have counterparts (i.~e.~points related to them by
$\star$) in the other spacetime and the others do not (corresponding
thus to singularities). It can be shown (see Lemma 1 in the Appendix)
that the first type points form a dense subset of $\Bd N$.
\par\medskip
In what follows we proceed from the assumption 
that the condition relating the two worlds is just
that they are described by spacetimes diverging by $S$ (with $N_i$
corresponding to the unchanged regions). It should be
noted, however, that this condition is  
tentative to some extent. It is not impossible that some other
conditions may be of interest, more restrictive than ours
(e.~g.~we could put some requirements on points 
of the second type) or, on the contrary,
less restrictive.
The latter can be obtained, for example, in the following
manner. The relation $\star$ is reflective and symmetric, but not
transitive. Denote by $\sim$ its transitive closure
(e.~g.~in the case depicted in Fig.~2, 
$A \;\lefteqn{\hbox to 1.2ex{\hfill/}}{\star}\; B$, but $A \sim B$).
\kart{!h,b,t}{30ex}{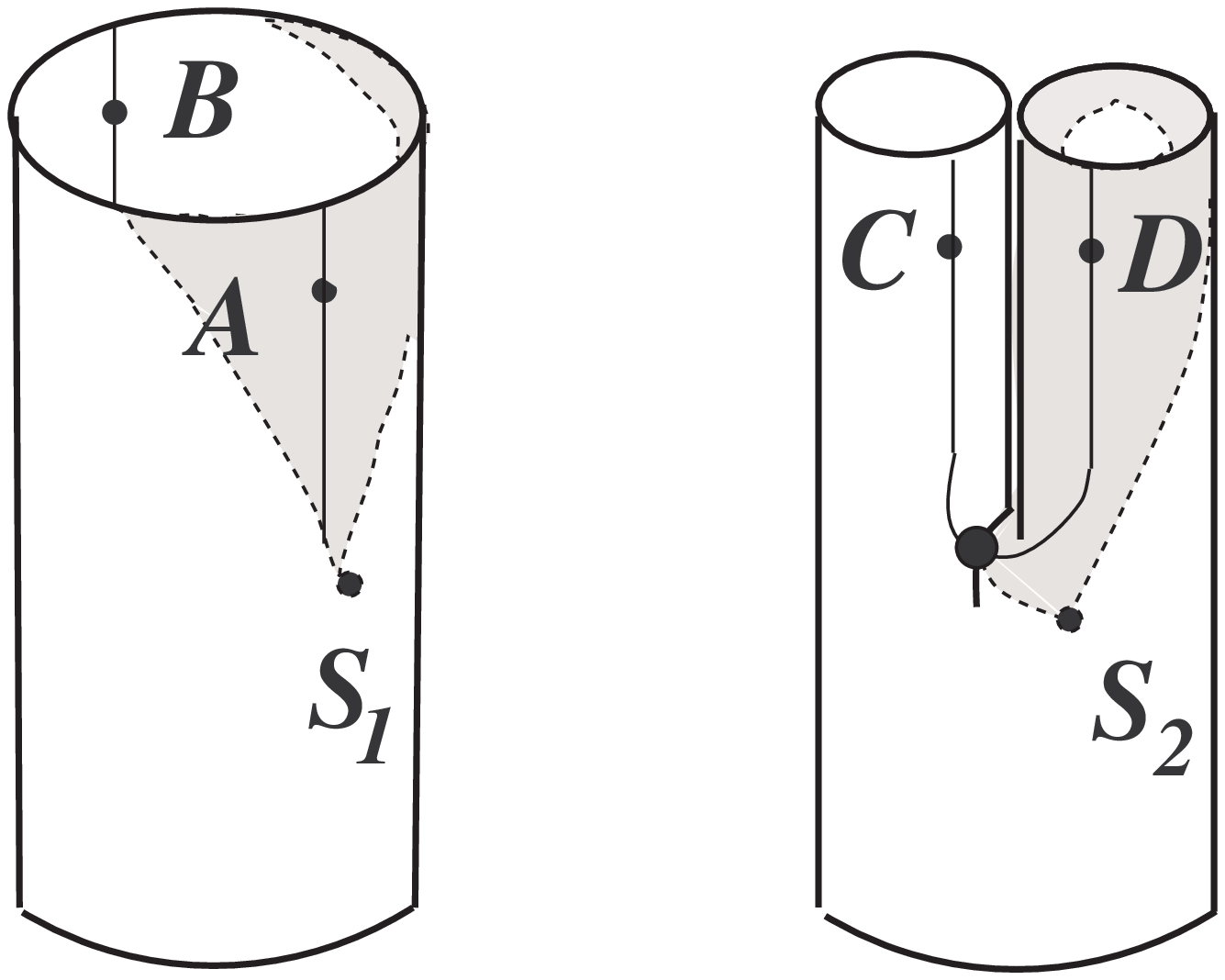}%
{Make cuts along the thin lines on the cylinder at the left and glue
  their banks to obtain the ``trousers'' at the right.  The shadowed
  regions depict $\JS{}$. Note that these spacetimes cannot be
  considered as diverging by $S$. For, if we take, for example, the whole
  $M_i$ with the thin lines removed, as $N_i$, then $\Bd N_1
  \ni B \sstar C$, while neither $B$ nor $C$ lies in $\mathbb J$.}
Now, if we want to consider topology changes like that in
Fig.~2 as possibly produced by the event $S$, we can replace
 (\ref{Con}) by the requirement that for any first type point $Q \in \Bd N_j$,
	\begin{equation}	
	[Q]_\sim \cap \mathbb J \neq \emptyset,
	\label{Con'}
	\end{equation}
where $[Q]_\sim \equiv \{x{|}\; x \sim Q\}$.
It is worth pointing out that replacing (\ref{Con}) by (\ref{Con'}) does
not actually affect any of the statements below. 
\par

Now we can formulate the question
posed in the Introduction as follows:\par\smallskip\noindent
\emph{Given that spacetimes $M_i$  diverged by an event $S$, 
how will the points $F_2,\:R_2$ be  related to the points $F_1,\:R_1$?}
\par\smallskip\noindent(It is understood from now on that
	$$
	\mathcal C_2 \cap N_2 = \phi(\mathcal C_1 \cap N_1),
	$$
where $\mathcal C_i = \mathcal{D}_i,\,\mathcal{E}_i$.)

\section{ One-way trip} 
Example 3 shows that contrary to what one might expect,  utter
causality by itself does not prevent  a pilot from
hastening the arrival at a destination.  It is reasonable to suppose,
however, that in less ``pathological'' spacetimes\footnote{%
Note that we discuss the causal structure only. So the fact that
there are singularities in the spacetime from Example 3 is 
irrelevant. As is shown in \cite{STM}, a singularity-free spacetime
can be constructed with the same causal structure.} this is not the case.

\paragraph{Proposition 1.} If $M_i$ are globally hyperbolic
spacetimes diverging by $S$, then
	$$
	F_1 \sstar F_2.
	$$

\par\smallskip The proof of this seemingly self-evident proposition
has turned out to be quite tedious, so we cite it in the Appendix.
\paragraph{Example 4.}                          
Recently it was proposed \cite{Alc} to use for
hyper-fast travel the  metric (I omit
 two irrelevant dimensions $y$ and $z$)
	\begin{equation}
        \label{Tah}                        
	ds^2=-dt^2+[dx-v_sf(r_s)dt]^2.
	\end{equation}
Here $r_s\equiv|x-x_s|,\ v_s(t)\equiv dx_s(t)/dt$,
and $x_s(t)$ and $f$
are arbitrary smooth  functions satisfying\footnote{ In         
\cite{Alc} another $f$ was actually used. Our modification,
however, in no way impairs the proposed spaceship.}
	$$
     x_s(t)=\cases{
		D & at  $t>T$ \cr
	        0 & at  $t<0$ \cr
		} 
	\qquad
	 f(\xi)=\cases{
               1 & for $\xi\in(-R+\delta,R-\delta)$\cr
               0 & for $\xi\notin(-R,R)$. \cr
		} 
         $$
$\delta,\ T$, and $R$ are arbitrary positive parameters.\par
%
To see the physical meaning of the condition of utter causality take the
Minkowski plane as $M_1$ and the plane endowed with the metric
(\ref{Tah}) as $M_2$ (the $S_i$ are meant to be the origins of the
coordinates). It is easy to see that the curve
$\lambda\equiv(t,x_s(t))$ is timelike with respect to
the metric (\ref{Tah}) for any $x_s(t)$. So we could conclude that an
astronaut can travel with an arbitrary velocity
(``velocity" here is taken to mean the coordinate
velocity $dx_a(t)/dt$, where $x_a(t)$ is the astronaut's
world line). All he needs is to choose an appropriate
$x_a(t)$ and to make the metric be of
form (\ref{Tah}) with $x_s(t)=x_a(t)$.
The distortion of the spacetime in the
region $\{0<x<D,\ t>0\}$ of $M_2$ will allow him to travel
faster than he could have done in the flat space $M_1$ (which
does not of course contradict the Proposition since the $M_i$ do not
diverge by $S$).
\par
The subtlety lies in the words ``to make the metric
be \dots." Consider the curve $\lambda_+\equiv (t,x_s(t)+ R)$, which
separates the flat and the curved regions.  It is easy to see that
$v_s(t)>1$ when and only when $\lambda_+(t)$ is spacelike.  At the same
time eq.~(19) of \cite{Alc} says that the space immediately to the
left of $\lambda_+$ is filled with some matter
($G^{00}\ne0$)\footnote{The case in point is, of course,
a four-dimensional space.}. The curve $\lambda_+(t)$ is thus the world
line of the leading edge of this matter. We come therefore to the
conclusion that to achieve $T<D$ the astronaut has to use tachyons.
This possibility is not too interesting: no wonder that one can
overcome the light barrier if one can use the tachyonic matter.
Alternatively,  in the more general case, when the spacetime is nonflat
from the outset, a similar result could be achieved without tachyons
by placing  {\em in advance} some devices along
the pilot's way and programming them to come into operation at preassigned
moments and to operate in a preassigned manner. Take the moment $P$ when
we began placing the devices as a point diverging the spacetimes.
Proposition~1 shows then that, though a regular spaceship
service perhaps can be set up by this means, it does not help to
outdistance the test particles from $M_1$ in the \emph{first} flight
(i.~e.~in the flight that would start at $P$).

\section{Round-trip} 
The situation with the points $R_i$ differs radically from that
with $F_i$ since the segment $FR\,$ belongs to $J^+(S)$ for sure. So
even in globally hyperbolic spacetimes there is nothing to prevent an
astronaut from modifying the metric so as to move $R$ closer to $S$
(note that from the viewpoint of possible applications to
interstellar expeditions this is far more important than to shift
$F$).  Let us consider two examples.

\paragraph{Example 5.  ``The warp drive."}         
Consider the metric
$$
     ds^2=-(dt-dx)(dt+k(t,x)dx),
$$
where $k\equiv 1-(2-\delta)\theta_\epsilon(t-x)
[\theta_\epsilon(x)-\theta_\epsilon(x+\epsilon-D)]$. Here
$\theta_\epsilon$ denotes a smooth monotone function: 
	$$
	\theta_\epsilon(\xi)=
	\cases{
	1 & at  $\xi>\epsilon$\cr
        0 & at  $\xi<0$ \cr
	} 
	$$
$\delta$ and $\epsilon<D$ being arbitrary small positive parameters. 
\par
\kart{h,t,b}{30ex}{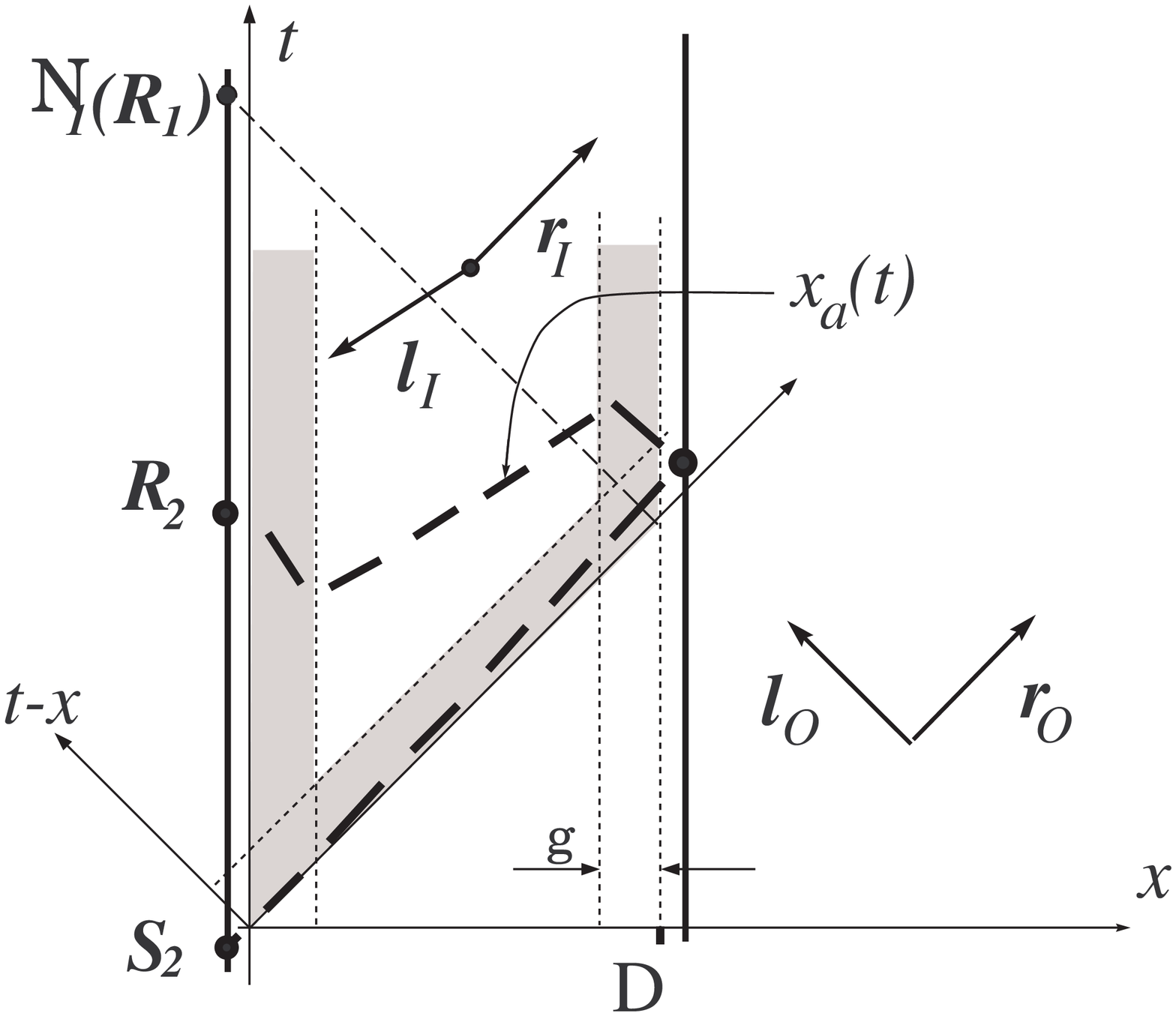}%
{Warp drive.}
\par\noindent
Three regions can be recognized in $M$ (see Fig.~3):
\par\smallskip\noindent
{\em The outside region:} $\{x<0\} \cup \{x>D\} \cup
\{x>t\}.$ The metric is flat here ($k=1$). Future
light cones are generated by vectors 
$\mathbf r_O=\partial_t+ \partial_x$ and 
$\mathbf l_O=\partial_t- \partial_x$ \\
{\em The transition region.} It is a narrow (of width
$\sim\epsilon$) strip shown as a shaded region in Fig.~3. 
The spacetime is curved here.\\
{\em The inside region:} $\{x<t-\epsilon\} \cap
\{\epsilon<x<D-\epsilon\}.$ This region is also
flat ($k=\delta-1$), but the light cones are ``more
open" here being generated 
by ${\mathbf r}_I=\partial_t+ \partial_x$ and 
${\mathbf l}_I=-(1-\delta)\partial_t- \partial_x$.
\par
The vector ${\mathbf l}_I$ is almost antiparallel to ${\mathbf r}_I$
and thus a photon moving from $F$ toward the left
will reach the line $x=0$ almost in $S$. 
\par
We see thus that an arbitrarily distant journey can be
made in an arbitrarily short time! It can
look like the following. In 2000, say, an astronaut ---
his world line is shown as a bold dashed line in Fig.~3
--- starts to Deneb. He moves with a near light speed
and the way to Deneb takes the (proper) time $\Delta \tau_a \ll 1600\,$yr
for him. On the way he carries out some manipulations
with the ballast or with the passing matter. In spite
of these manipulations the traveller reaches Deneb at
3600 only. However, on his way back he finds that the
metric has changed and he moves ``backward in time,"
that is, $t$  decreases as he approaches the Earth
(though his trajectory, of course, is {\em future-directed}).
As a result, he returns to the Earth in 2002.

\paragraph{Example 6. Wormhole.} 
Yet another way to return arbitrarily soon after the
start by changing geometry is the use of wormholes.
Assume that we have a wormhole with a negligibly
short throat and with  both mouths resting {\em near the
Earth.} Assume further that we can move any mouth at
will without changing the ``inner" geometry of the
wormhole. Let the astronaut take one of the mouths with
him. If he moves with a near light speed, the trip
will take only the short time $\Delta\tau_a $ for him.
According to our assumptions the clocks on the Earth as
seen through the throat will remain synchronized with
his and the throat will remain negligibly short.
So, if immediately after reaching Deneb he returns to
the Earth through the wormhole's throat, it will turn
out that he will have returned within 
$\Delta \tau_\mathcal E \approx \Delta \tau_a $ after the start.
\par
Similar things were discussed many times in connection
with the worm\-hole-based time machine. The main
technical difference between a time machine and a
vehicle under consideration is that in the latter case the mouth
only moves \emph{away} from the Earth. So causality is
preserved and no difficulties arise connected with 
its violation. 
\par

\section{Discussion}

In all examples considered above the pilot, roughly speaking,
``transforms'' an ``initially'' spacelike (or even past-directed)
curve into future-directed.  Assume now that one applies this
procedure first to a spacelike curve $(AC_1B)$ and then to another
spacelike curve $(BC_2A)$ lying in the intact until then region. As a
result one obtains a closed timelike curve $(AC_1BC_2A)$ (see
\cite{AE,EveRo,Tho} for more
details).  So the vehicles in discussion can be in a sense
considered as ``square roots'' of time machine (and thus a collective
name \emph{space machine} --- also borrowed from science fiction ---
seems most appropriate for them). The connection between time and
space machines allows us to classify the latter under two types.\\
{\bf 1.} Those leading to time machines with compactly generated
Cauchy horizons (Examples 4--6). From the results of \cite{Conj} it
is clear that the creation of a space machine of this type requires
violation of the weak energy condition. The possibility of such
violations is restricted by the so-called ``quantum inequalities'', QIs
\cite{QI}.
In particular, with the use of  a QI it was shown in \cite{EveRo} that
to create a four-dimensional analog of our Example 5 one needs huge
amounts (e.~g.~$10^{32} M_{galaxy}$) of ``negative energy".
Thermodynamical considerations suggest that this in its turn
necessitates huge amounts of ``usual" energy, which makes the creation
unlikely. This conclusion is quite sensitive to the details of the geometry
of the space machine and one could try to modify its construction so
as to obtain 
more appropriate values. Another way, however, seems more promising.
The QI used in \cite{EveRo} was derived with the constraint (see
\cite{QI}) that in a 
region with the radius smaller than the proper radius of curvature spacetime
is ``approximately Minkowski'' in the sense that the energy density
(to be more precise, the integral 
$E[\lambda,\tau_0,T] \equiv \int_{-\infty}^{T}
 \langle T_{\mu\nu} u^{\mu} u^{\nu} \rangle
({\tau}^2+{\tau_0}^2)^{-1} \, d\tau$, where $\lambda$ is a timelike
geodesic parametrized by the proper time $\tau$, $\mathbf u \equiv
\partial_\tau$, and $\tau_0$ is a ``sampling time")
is given by essentially the same expression as in the Minkowski
space.  So, in designing space machines,
spacetimes are worth searching for where this constraint breaks down.
\par
Among them is a ``critical'' (i.~e.~just before its transformation
into a time machine) wormhole. Particles propagating through such a
wormhole again and again experience (regardless of specific
properties of the wormhole \cite{Cl}) an increasing blueshift. The
terms in the stress-energy tensor associated with nontrivial
topology also experience this blue-shift \cite{Yur}. As a result, in
the vicinity of the Cauchy horizon (even when a region we consider is
flat and is located far from either mouth) the behaviour of the
energy density has nothing to do with what one could expect from the
``almost Minkowski'' approximation \cite{Inst}. (The difference is so
great that \emph{beyond} the horizon we cannot use the
known quantum field theory, including its methods of evaluating the
energy density, at all \cite{Wal}.) Consider, for example, the Misner
space with the massless scalar field in the conformal vacuum state.
From the results of Sec.~III.B \cite{Inst} it is easy to see that
$E[\lambda,\tau_0,\infty] = -\infty$ for \emph{any} $\lambda$ and
$\tau_0$ and the QI thus does not hold here\footnote{It is most
likely (see Sec.~IV of \cite{Conj}) that the same is true in the
four-dimensional Misner space  as well.}. Moreover,
$E[\lambda,\tau_0,T] \to -\infty$ as one approaches the Cauchy
horizon along $\lambda$. So, we need not actually create a time
machine to violate the QI. It would suffice to ``almost create" it.
\par
Thus it well may be that in spite of (or owing to) the use of a
wormhole the space machine considered in  Example 6 will turn out
to be more realistic than that in Example 5.\par\noindent
{\bf 2.} Noncompact space machines, as in Example 3. These (even their
singularity free versions; see \cite{STM,Me}) do not necessitate
violations of the weak energy condition. They have, however, another
drawback typical for time machines. The evolution of nonglobally
hyperbolic spacetimes is not understood clearly enough and so we do not
know how to \emph{force} a spacetime to evolve in the appropriate way.
There is an example, however (the wormhole-based
time machine \cite{TM}), where the spacetime is denuded of its global
hyperbolicity by quite conceivable manipulations, which gives us some
hope that this drawback is actually not fatal.

\section*{Acknowledgments}
This work is partially supported by
the RFFI grant 96-02-19528. I am grateful to D.~Coule, A.~A.~Grib,
G.~N.~Parfionov, and R.~R.~Zapatrin 
for useful discussion.

\appendix
\section*{Appendix}
Throughout this section we take $M_i$ to be globally hyperbolic
spacetimes diverging by $S$ and $\BS{U}$ to mean $U \setminus \mathbb
J$ for any set $U$.

\paragraph{Lemma 1.} Let $O$ be a neighbourhood of a point of 
$\Bd N_j$ and ${O^N}\subset O\cap N_j$ be such an open nonempty set that
	\begin{equation}
	\Bd {O^N}\cap O\subset \Bd N_j
	\label{L11}
	\end{equation}
Then
	$$
	\Bd {O^N}_{(i)}\cap \JS{i}\neq\emptyset\qquad
	\mbox{for some }i.
	$$

\paragraph{\it Proof.} Let $j = 1$ for definiteness. Consider a
smooth manifold $\widetilde{M}\equiv M_2\cup_{\phi'}O$, where
${\phi'}$ is the restriction of ${\phi}$ on ${O^N}$. Induce the
metric on $\widetilde{M}$ by the natural projections 
	$$
	\pi_i\colon\qquad M_2\stackrel{\pi_1}{\longmapsto}\widetilde{M},
	\quad O\stackrel{\pi_2}{\longmapsto}\widetilde{M}
	$$
(or, more precisely by $\pi_i^{-1}$) thus making  $\widetilde{M}$ into
a Lorentzian manifold and $\pi_i$ into  isometrical embeddings.
$\widetilde{M}$ must be non-Hausdorff since otherwise it would be a
spacetime and so (as $M_2\varsubsetneq\widetilde{M}$) $M_2$ would have
an extension in contradiction to its definition. So points $Q_i$
exist:
	\begin{equation}
	Q_1\sstar Q_2,\qquad Q_1\in \Bd {O^N}_{(1)}\cap O,
	\quad Q_2\in \Bd {O^N}_{(2)}
	\label{L12}
	\end{equation}
and the lemma follows now from Def.~1 coupled with
(\ref{L11}). 
\par\nopagebreak\hfill$\square$

\paragraph{Lemma 2.} If both $A_{(i)}$ lie in $\BS{N_i}$, then
so do $I^-(A_{(i)})$.

\paragraph{\it Proof.} $M_i$ are globally hyperbolic. So any point
$P$ has such a neighbourhood  (we shall denote it by $V_P$) that,
first, it is causally convex,
i.~e.~ $ J^-(x) \cap J^+(y) \subset V_P $ for any points
$x,\,y \colon\; y \in J^-(x,V_P) $; and, second, it lies in a convex normal
neighbourhood of $P$. Now suppose the lemma were false. We could
find then such a point $A' \in I^-(A_{(i)},N_i)$ (let $i=1$, for
definiteness) that
	$$ 
	W  \neq I^-(A', V_{A'}),
	$$
where $ W \equiv I^-(A',N_1\cap V_{A'})$.\par
\kart{h,t,b}{30ex}{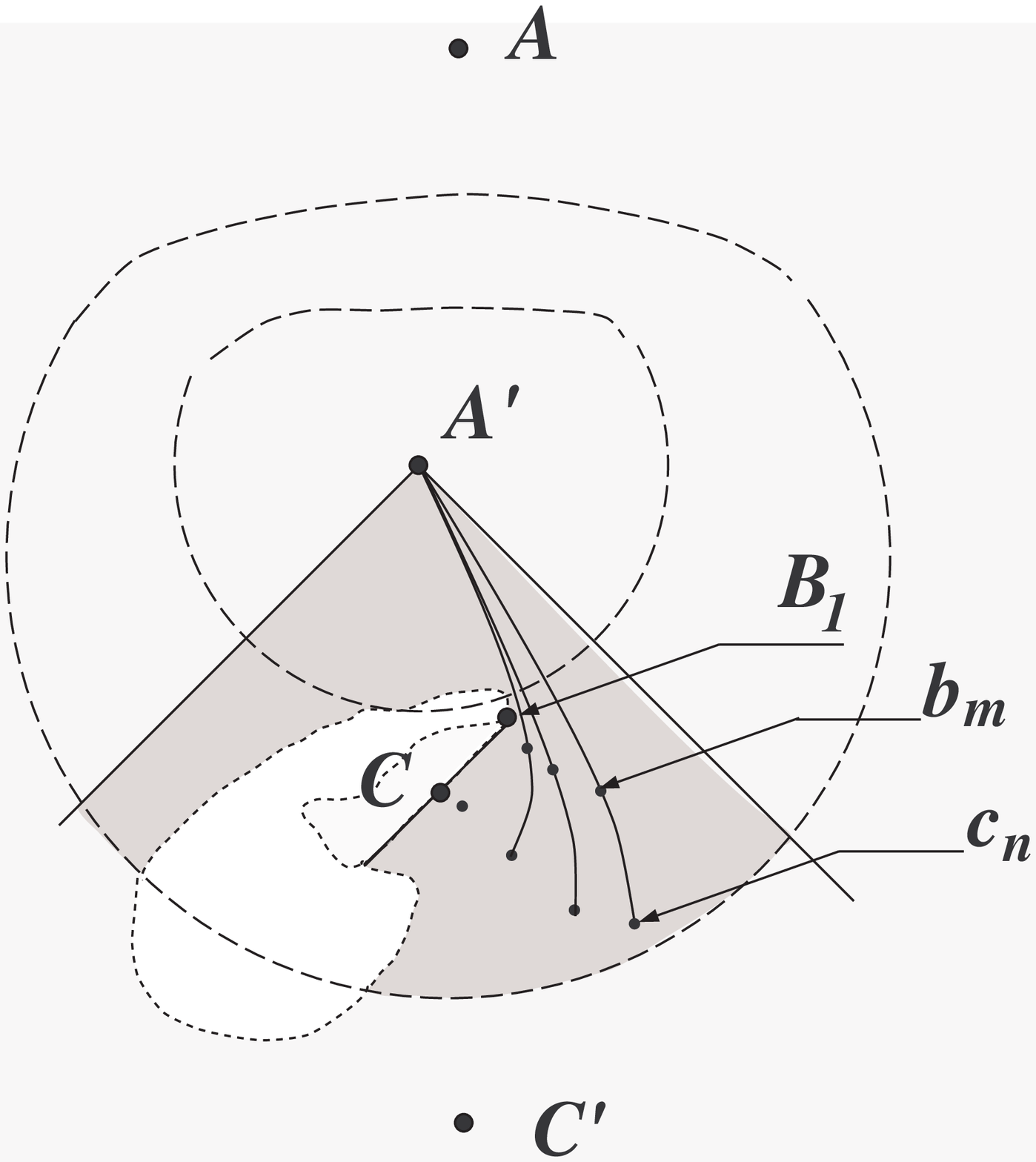}%
{Case 1 of Lemma 2. The white area does not belong to $N_1$,
the darkest area is 
$W$. If instead of the larger area bounded by a dashed line we
take the smaller one as $V_{A'}$, we get Case 2.}
\noindent Denote $\Bd W \cap I^-(A',V_{A'})$ by $\partial W$. Clearly 
$\emptyset \neq \partial W  \subset \overline{N_1}$.
So, let us consider the two possible cases (see Fig.~4):\\
\textbf{I.} $\partial W \not\subset \Bd N_1$.\\
Under this condition a point $C$ and a sequence of causal curves
$\{\gamma _n\}$ from $A'$ to points $c_n$ exist such that
	$$
	\gamma_n \subset W,\quad
	c_n \to C \in \partial W \cap N_1
	$$
According to \cite[Prop.~2.19]{BeE} there exists a causal curve
$\gamma$ connecting $A'$ and $C$, which is limit for $\{\gamma _n\}$
and is lying thus in $ \overline{W}$.
Since $V_{A'}$ belongs to a normal convex neighbourhood and $C\in
I^-(A',V_{A'})$, 
$\gamma$ by \cite[Prop.~4.5.1]{HawE} is not a null geodesic and hence
	\begin{equation}
	\gamma\not\subset N_1
	\label{L21}
	\end{equation}
(otherwise by \cite[Prop.~4.5.10]{HawE} and by causal convexity of
$V_{A'}$ we could deform it into a timelike 
curve lying in $N_1  \cap V_{A'}$, while $C\notin W$).
\par
Now note that for any $C'\in I^-(C,N_1)$ there exists a subsequence
$\{\gamma _k\}$ lying in $I^+(C',N_1) $. So by 
(\ref{L21}) a sequence of points $\{b_m\}$ and a point $B_1$ can be
found such that
	\begin{equation}
	b_m\to B_1\in \Bd N_1,
	\qquad b_m\in I^-(A',N_1)\cap I^+(C',N_1) .
	\label{L22}
 	\end{equation}
Thus the $\phi(b_m)$ lie in a compact set 
$J^-(A'_{(2)})\cap J^+(C'_{(2)})$ and therefore
	$$
	\phi(b_m)\to B_2\colon\quad B_1\sstar B_2.
	$$
From Def.~1 it follows that at least one of the $B_i$ lies in \JS{i} and
since $B_i\in I^-(A_{(i)})$ we arrive at a contradiction.
\\
\textbf{II.} $\partial W \subset \Bd N_1$.\\
In this case taking $O=I^-(A',V_{A'})$ and ${O^N}=W$ in Lemma 1 yields
	$$
	\overline{W_{(i)}} \cap \JS{i} \neq \emptyset
	\qquad  \mbox{for some }i,
	$$
which gives a contradiction again since $\overline{W_{(i)}} \subset
I^-(A_{(i)})$. 
\par\nopagebreak\hfill$\square$\par
\noindent
Consider now the sets $L_i \equiv \{x|\; I^-(x) \subset N_i\}$. They
have a few obvious features:
	\begin{eqnarray}
	L_i = \overline{\Vn L_i}, \quad 
	\Vn L_i \subset N_i  \label{L32}
	\\
	A_{(1)} \in \BS{L_1}
	\quad \Leftrightarrow \quad
	A_{(2)} \in \BS{L_2}
	\label{L33}
	\end{eqnarray}
Combining Lemma 2  with (\ref{L32},\ref{L33}) we obtain:
	\begin{equation}
	\BS{ \Bd L_i } \subset \Bd N_i .
	\label{L34}
	\end{equation}

\paragraph{Lemma 3.} $ \BS{L_i} = \BS{M_i} $.
\paragraph{\it Proof.} Since $ \BS{M_i} $ is connected and $\BS{ \Vn
L_i }$ is non-empty [e.~g.~from Def.~1 $I^-(S_i) \subset \BS{ \Vn
L_i }$] it clearly
suffices to prove that  $ \BS{\Bd L_i} = \emptyset $.
To obtain a contradiction, suppose  that there exists a point $A \in
\BS{\Bd L_1} $ 
 and let $U$ be such a neighbourhood of $A$ that
	$$
	\overline U \subset \BS{M_1} 
	$$
Then for ${U^L} \equiv U \cap \Vn L_1$ it holds that
	$$
	\overline{{U^L}_{(i)}} \cap \JS{i} = \emptyset 
	\qquad i=1,2.
	$$
On the other hand, owing to (\ref{L32},\ref{L34}) we can take $O=U$
and ${O^N}={U^L}$ in Lemma~1 and get 
	$$
	\overline{{U^L}_{(i)}} \cap \JS{i} \neq \emptyset 
	\qquad  \mbox{for some }i.
	$$
Contradiction.
\par\nopagebreak\hfill$\square$\par
\paragraph{Corollary 1.} 
$ \BS{I^+(\mathcal{E}_2)} =   \phi\big( \BS{ I^+(\mathcal{E}_1) }\big) $.
\subsubsection*{ Proof of Proposition 1.} $M_i$ is causally simple.
Hence a segment of null geodesic from $S_i$ to $F_i$ exists. By
\cite[Prop.~4.5.10]{HawE} this implies that any point 
$P_{(i)} \in \BS{\mathcal{E}_i}$ 
can be connected to $F_i$ by a timelike curve. Hence a point
$P'\in \BS{\mathcal{D}_i}$ can be reached from $P_{(i)}$ by a timelike
curve without intersecting $\JS{i}$.
Thus $F_i$ is the future end point of the curve
$\mathcal{D}'_i$: 
	$$
	\mathcal{D}'_i \equiv \mathcal{D} \cap \BS{ I^+(\mathcal{E}_i) }
	$$
And from Corollary 1 it follows that $\phi( \mathcal{D}'_1 ) = 
\mathcal{D}'_2 $.
\par\nopagebreak\hfill$\square$\par

\end{document}